\documentclass[aip,reprint,amssymb,amsmath,superscriptaddress,longbibliography]{revtex4-1}

\usepackage{graphicx}
\usepackage{dcolumn}
\usepackage[T1]{fontenc}
\usepackage{lmodern}
\usepackage{hyperref}
\preprint{AIP/123-QED}

\usepackage{xcolor}
\usepackage[caption=false]{subfig}
\DeclareMathOperator{\tr}{tr}

\usepackage{epsfig}
\usepackage{epstopdf}

\begin{document}
\title{Nonlinear $q$-voter model from the quenched perspective}
\author{Arkadiusz J\k{e}drzejewski} 
\affiliation{Department of Theoretical Physics, Wroc\l{}aw University of Science and Technology, 50-370 Wroc\l{}aw, Poland}
\author{Katarzyna Sznajd-Weron} 
\affiliation{Department of Theoretical Physics, Wroc\l{}aw University of Science and Technology, 50-370 Wroc\l{}aw, Poland}
\date{\today}

\begin{abstract}

We compare two versions of the nonlinear $q$-voter model: the original one, with annealed randomness, and the modified one, with quenched randomness.
In the original model, each voter changes its opinion with a certain probability $\epsilon$ if the group of influence is not unanimous.
In contrast, the modified version introduces two types of voters that act in a deterministic way in case of disagreement in the influence group: the fraction $\epsilon$ of voters always change their current opinion, whereas the rest of them always maintain it.
Although both concepts of randomness  lead to the same average number of opinion changes in the system on the microscopic level, they cause qualitatively distinct results on the macroscopic level. 
We focus on the mean-field description of these models. Our approach relies on the stability analysis by the linearization technique developed within dynamical system theory. This approach allows us to derive complete, exact phase diagrams for both models. The results obtained in this paper indicate that quenched randomness promotes continuous phase transitions to a greater extent, whereas annealed randomness favors discontinuous ones. The quenched model also creates combinations of continuous and discontinuous phase transitions unobserved in the annealed model, in which the up-down symmetry may be spontaneously broken inside or outside the hysteresis loop. The analytical results are confirmed by Monte Carlo simulations carried out on a complete graph.
\end{abstract}
\maketitle
% Tak jak chcieli jest blisko 250 slow

\textbf{The subject of the opinion dynamics is studied within social and computer science, mathematics, statistical physics, and engineering, and models of binary opinions are of particular interest within all these disciplines. Among them the $q$-voter model is one of the most general one, reducing for specific values of parameters to  other models, including the famous voter model. In the original $q$-voter model all agents are identical and can change randomly their behavior in time, which corresponds to the so-called annealed approach. 
In this paper, we reformulate the model under the quenched approach, i.e., we assign to agents some individual traits that remain constant in time, and ask the question about the role of the approach in shaping macroscopic behavior of the model. 
To answer the question, we compare two versions of the model, the quenched and the annealed one. Such a comparison is an important issue in statistical physics because the annealed approach is much simpler for the analytical treatment, and it is often used as an approximation of the real quenched system. Moreover, it may be also interesting from the social point of view because it corresponds to the famous \textit{person-situation debate}.}

\section{Introduction}
Randomness plays a leading role in modeling complex systems. 
It is not only crucial for capturing characteristic structural properties of real-world networks \cite{Watt:Str:98,Bar:Alb:99,Alb:Bar:02}, which are further used as underlying settings for various dynamics, but also many dynamics themselves hinge on random processes \cite{Gle:13,Jed:Szn:19,Red:19}.
In statistical mechanics, one can come across two specific types of randomness -- annealed and quenched.
The former is related with some properties of a system that can vary randomly in time.
In contrast, features that do not change with system progression although they are randomly distributed at the beginning of the process are associated with the latter 
\cite{Hin:00,Ste:New:13}.
In particular, annealed and quenched randomness are used to model different kinds of disorders and irregularities present in nature. 
Thus, comprehending their impact on a system is of great importance.
It is not surprising, then, that these two concepts are already well established in condensed matter physics.
Thermal fluctuations of atoms, for instance, are manifestations of annealed randomness, whereas spin glasses, models of disordered magnetic materials, which date from the mid-1970s, are inherently connected with quenched randomness \cite{Ste:New:13}.
In fact, the theoretical success of spin glasses quickly sparked the cascade of works incorporating the same concepts.
Annealed and quenched randomness began to be related with interactions between spins \cite{Jav:Mar:18,Tho:Bee:76,Lee:etal:09,Mal:Val:Das:14,Lip:Gon:Lip:15,Jed:Chm:Szn:17,Par:Noh:17,Edw:And:75,Ber:etal:17,Aiz:Weh:89,Hui:Ber:89,Cha:Ber:98} or with external fields \cite{Ber:etal:17,Aiz:Weh:89,Hui:Ber:89}.
Consequently, different disorders have been studied in many spin models, including Ising model \cite{Tho:Bee:76,Lee:etal:09,Mal:Val:Das:14,Lip:Gon:Lip:15,Jed:Chm:Szn:17,Par:Noh:17}, Edwards-Anderson model \cite{Edw:And:75}, or XY model \cite{Ber:etal:17}.
Along with the development of network science,  these two approaches to modeling random events have come into the spotlight again.
Adaptive and static network models are  good examples of systems with annealed and quenched disorders, respectively \cite{Boc:etal:06,Bol:etal:12}.

However, different kinds of randomness are not only relevant to natural sciences.
Human behavior can be also viewed through the prism of these ideas.
Such a relation becomes especially interesting from the perspective of modeling social systems and opinion dynamics \cite{Jed:Szn:19}.
Let us bring up two opposing psychological theories that can be directly associated with quenched and annealed randomness.
In fact, they are the subject of the long-standing person-situation debate, sparked off already in the late 1960s \cite{Ken:Fun:88} -- interestingly, this date almost coincides with a boom in research on spin glasses.
It turns out that some psychologists believe that people are characterized by a set of stable and enduring qualities. 
These dispositions define their personality and distinguish them from others \cite{Ken:Fun:88,Kra:92,Fun:08}.
Consequently, human behavior, which reflects one's personality, should also be persistent and predictable, at least to some extent.
Such a point of view on human behavior, as a personal feature that does not change in time although varies between individuals in the system, fits in with the definition of quenched randomness.
However, next to personality psychologists, which are in favor of this ideology, there are social psychologists, which take the opposite viewpoint on this issue.
They regard behavior as a response to situational factors rather than a manifestation of personal dispositions \cite{Ken:Fun:88,Kra:92,Fun:08}.
This in turn resembles more the concept of annealed randomness since variable conditions of different situations may lead to the behavior that changes over time.
Although it seems that eventually both sides of the debate have their points, and human behavior results from the interplay between dispositions and situations \cite{Fun:08}, testing these extreme ideas in models of opinion dynamics is still tempting.

In fact, the concept of quenched randomness in the form of individuals with fixed characteristic behaviors in the system has already appeared in studies on the opinion formation.
We can find these individuals under different names in works on the voter \cite{Jav:Squ:15,Tan:Mas:13,Jed:Szn:17,Kha:Tor:19,Mau:03,Mob:15,Mel:Mob:Zia:17,Kha:San:Tor:18,Mel:Mob:Zia:16}, majority-vote \cite{Jav:14,Vil:etal:18}, Galam \cite{Sta:Mar:04}, or Sznajd model \cite{Sch:04}.
Next to conformists, which tend to follow the crowd and show up in all listed references, we can encounter anticonformists \cite{Jav:14,Jav:Squ:15,Jar:etal:15,Jed:Szn:17}, called also contrarians \cite{Tan:Mas:13,Sta:Mar:04,Vil:etal:18,Kha:Tor:19,Sch:04}, which rebel against social norms, or independent individuals \cite{Nyc:Szn:Cis:12,Jed:Szn:17}, which make their own decisions regardless of social pressure. 
Zealots, on the other hand, favor one opinion to the extent that they can independently switch to it \cite{Mau:03}.
Their beliefs become even stronger in more recent studies on voter models where they do not change the opinions at all \cite{Mob:15,Mel:Mob:Zia:17,Kha:Tor:19, Kha:San:Tor:18,Mel:Mob:Zia:16}.
Similar adamant individuals can be found in the work on the Sznajd model \cite{Sch:04}.

The analyses of quenched and annealed randomness alone contributes undoubtedly to the development of social agent-based models.
However, comparative studies where these approaches are directly confronted may provide additional information on phase transitions.
In some models, quenched randomness tends to shift the transition point of continuous phase transitions in such a way that the ordered phase becomes wider.
The comparison of Galam's original model \cite{Gal:04}, introduced in the spirit of annealed randomness, and its quenched version \cite{Sta:Mar:04} leads to this conclusion.
The same applies to the $q$-voter model with independence \cite{Jed:Szn:17,Nyc:Szn:Cis:12}.
However, this effect is not universal. 
The $q$-voter model with anticonformity displays exactly the same continuous phase transitions under both, quenched and annealed, approaches \cite{Jed:Szn:17}.
How the randomness impacts the phase transition type is another interesting issue.
In the $q$-voter model with independence, discontinuous transitions are also possible \cite{Nyc:Szn:Cis:12}.
After transforming the model to the quenched version, continuous phase transitions take the place of all discontinuous ones \cite{Jed:Szn:17}.
This may suggest that the quenched approach favors continuous phase transitions over discontinuous ones as in the equilibrium statistical physics \cite{Aiz:Weh:89,Hui:Ber:89,Cha:Ber:98}.
Certainly, more comparative studies are required to verify these conjectures.
Our manuscript extends the literature in this regard by contrasting two formulations, annealed and quenched, of the original nonlinear $q$-voter model without additional nonconformity.

The nonlinear $q$-voter model \cite{Cas:Mun:Pat:09} is one of many models of opinion dynamics, which originally implements the concept of annealed randomness \cite{Gle:13,Jed:Szn:19}.
The process takes place on a network that represents a social structure. 
Nodes correspond to voters, whereas links mark some kind of relation between them.
Each voter holds an opinion in a form of a two-state variable $j\in\{1,-1\}$.
In such a setting, one voter after another is selected at random, and it interacts with its $q$ randomly chosen neighbors.
If all of them share the same opinion, the voter conforms and adjusts its opinion to the group.
However, if the selected neighbors are not unanimous, the voter changes its current state to the opposite one with probability $\epsilon$.
It means that the behavior of voters in case of disagreement in the influence group is not deterministic and may change in time.
Consequently, we deal with the annealed randomness here, and we will refer to the original version of the model \cite{Cas:Mun:Pat:09} as the annealed formulation of the model or the annealed model for brevity. 
From the psychological point of view, this kind of the probabilistic variation in the behavior of voters could be interpreted as a response to variable situational factors.
At this point, the question about quenched randomness naturally arises.
What would happen if the voters had personal dispositions, inclinations to stick with or give up their current opinions?
Would we see any changes on the macroscopic level? 
Would other kinds of phase transitions appear?
In this comparative study, inspired by the person-situation debate and research on spin glasses, we are going to look at the original $q$-voter model from the quenched perspective and answer the above questions.

\section{Quenched formulation of the model}
\label{sec:model}
In the quenched approach, the population of voters is divided into two fractions.
The first one gathers the vacillating individuals that always change their opinions when they run into a divided influence group.
The unyielding voters that stick with their viewpoints in such cases form the second fraction.
Let us stress that an unyielding voter can also change its opinion, but this happens only when the group of influence is unanimous.
In fact, Solomon E. Asch, a pioneering social psychologist, came up with evidence to suggest that unanimous groups of influence are much more persuasive \cite{Asc:51}. 
Thus, the behavior of unyielding voters only reflects this observation.
Similar idea of preserving the current opinion in case of a divided influence group shows up in a study on the Sznajd model \cite{Szn:Szn:00}.
To facilitate the comparison between the original $q$-voter model and our quenched modification, the vacillating group constitutes a fraction $\epsilon$ of the entire population of voters, whereas the remaining part $1-\epsilon$ falls into the unyielding group.
This ensures, on average, the same number of opinion changes in case of interactions with a divided influence group at the same level of the parameter $\epsilon$ in both approaches.

In summary, the quenched dynamics boils down to the following steps:
\\\\
\textit{\underline{Initiate the system:}}
\begin{enumerate}
	\item \textit{Assign dispositions to the voters.} Iterate over nodes in the network. A node is assigned to the vacillating group with the probability $\epsilon$ or to the unyielding one with the complementary probability $1-\epsilon$.
	\item \textit{Assign initial opinions to the voters.} In general, various proportions of voters with opposing opinions can be considered. We study populations in which these opinions are initially divided equally among voters. Thus, iterate over nodes in the network, and set a node's opinion to $1$ with probability $1/2$ or to $-1$ otherwise.
\end{enumerate}	

\noindent\textit{\underline{Run in a loop:}}
\begin{enumerate}
	\item \textit{Select at random one node in the network.} It represents a voter that is to reconsider its opinion.
	\item \textit{Choose randomly the members of the influence group.} Select at random $q$ neighbors of the voter. As in the original model \cite{Cas:Mun:Pat:09}, we allow repetition.
	\item \textit{Subject the voter to the social influence.} If all the group members hold the same opinion, the voter yields to the social pressure and adjust its opinion to the group. If the group is divided, the result depends on the voter's inclination:
	\begin{enumerate}
		\item A vacillating voter changes its opinion to the opposite, whereas
		\item an unyielding voter stays with its current opinion.
	\end{enumerate}
\end{enumerate}
In such a formulation of the model, $q$ is a positive integer.
However, the possible values of the parameter $q$ can be extended to all positive real numbers similarly as in Ref.~\cite{Cas:Mun:Pat:09}.

\section{Mean-field analysis and Monte Carlo simulations}
In this section, we are going to compare the annealed and quenched formulations of the nonlinear $q$-voter model.
Let us stress that the annealed formulation of the model corresponds exactly to the original model introduced in Ref.~\cite{Cas:Mun:Pat:09}.
This model version has already been analyzed on different structures \cite{Jed:Szn:19}, including regular latices and complete graphs \cite{Cas:Mun:Pat:09} or random regular networks \cite{Mor:etal:13}. 
Thus, Sec.~\ref{sec:annealed-model} only summarizes the behavior of the annealed model, whereas Sec.~\ref{sec:quenched-model} is devoted to the detailed analysis of the quenched model, introduced in Sec.~\ref{sec:model}.

We focus on the mean-field description of the models, which means that we consider them on an infinite complete graph so that all voters may interact with each other. 
Such a system where all its components may interact mutually is called well-mixed \cite{Jed:Szn:19}.
Of course, our Monte Carlo simulations are carried out for finite but very large well-mixed systems (i.e., with $N=5\cdot10^5$ voters).
The annealed model was already analyzed on the same structure in Ref.~\cite{Cas:Mun:Pat:09}.
However, in this reference, the exact potential of the system is approximated by the first terms of its power series expansion with additional constraints accounting for two absorbing states of the model \cite{Ham:etal:05,Vaz:Lop:08}.
As a result, the authors obtained only approximate phase diagram for the analyzed system.
We do not use such an approximation. 
Our analytical approach relies on the stability analysis by the linearization technique developed within dynamical system theory \cite{Str:94}, and it corresponds to the study of the exact potential.
Thus, Sec.~\ref{sec:annealed-model} not only outlines but also refines the mean-field description of the nonlinear $q$-voter model. Moreover,
the differences between the approximate and our exact results, confirmed by the Monte Carlo simulations, are highlighted at the end of this section.

In our study, we are particularly interested in the relation between the stationary values of the concentration of voters with opinion $j=1$, denoted by $c$, and the model parameters $\epsilon$ and $q$.
Throughout the work, we distinguish between the following phases: 
\begin{itemize}
	\item A fully ordered phase if $c\in\{0,1\}$ -- this phase corresponds to the consensus in the system. This state is dynamically inactive: according to the definition of the model, the dynamics freezes at this state, i.e., further changes in the voters' opinion are impossible. Such states that can be reached but cannot be left by the dynamics are called absorbing states. They cannot obey detailed balance with any active state, so the system with absorbing states is by definition out of equilibrium \cite{Hin:00}.
	\item A disordered phase if $c=0.5$ -- this phase corresponds to the situation where both opinions are equally likely in the system.  This state is dynamically active.
	\item An active ordered phase if $c\not\in\{0,0.5,1\}$ -- this phase covers all the remaining cases, and it corresponds to the system in which one opinion dominates over the other, but the consensus is not reached. In contrast to the fully ordered phase, this state is dynamically active from the microscopic point of view.
\end{itemize}

\subsection{Annealed model}
\label{sec:annealed-model}
The dynamics of our model in the mean-field approximation is given by the rate equation
\begin{equation}
\frac{dc}{dt}=F_\epsilon,
\label{eq:rateeqA}
\end{equation}
where $F_\epsilon$ can be interpreted as an effective force acting on the system \cite{Jed:Szn:19}.
Under the annealed approach, this force has the following form
\begin{equation}
F_\epsilon=(1-c)f_\epsilon(c)-cf_\epsilon(1-c),
\label{eq:FannealedGen}
\end{equation}
and it is derived based on the probability that a voter changes its opinion surrounded by the fraction $x$ of disagreeing neighbors 
\begin{equation}
	f_\epsilon(x)=x^q+\epsilon\left[1-x^q-(1-x)^q\right].
	\label{eg:flipprob}
\end{equation}
The above function is the same for all the voters in case of annealed randomness \cite{Cas:Mun:Pat:09}.
For $q=1$, the opinion change probability $f_\epsilon(x)=x$, so the original voter model is recovered.
Combining Eqs.~(\ref{eq:FannealedGen}) and (\ref{eg:flipprob}), we get
\begin{align}
F_\epsilon=&(1-c)c^q-c(1-c)^q\nonumber\\
&+\epsilon(1-2c)\left[1-c^q-(1-c)^q\right].
\label{eq:Fannealed}
\end{align}
Steady points of Eq.~(\ref{eq:rateeqA}) are those for which
\begin{equation}
F_\epsilon=0.
\label{eq:stCon}
\end{equation}
Next to the obvious solutions $c_\text{st}=\{0, 0.5,1\}$ for which Eq.~(\ref{eq:stCon}) is fulfilled for arbitrary values of $\epsilon$, we have the dependency
\begin{equation}
	\epsilon_\text{st}=\frac{c_\text{st}(1-c_\text{st})^q-(1-c_\text{st})c_\text{st}^q}{(1-2c_\text{st})\left[1-c_\text{st}^q-(1-c_\text{st})^q\right]}
	\label{eq:mstatesA}
\end{equation}
for the remaining solutions.
The sign of the derivative of the force with respect to the concentration at a steady point
\begin{equation}
\left.F_\epsilon'(c_\text{st})=\frac{dF_\epsilon(c)}{dc}\right|_{c=c_\text{st}}
\end{equation}
carries information about its stability. The state is stable if the result is negative $F'(c_\text{st})<0$, and unstable if positive $F'(c_\text{st})>0$ \cite{Str:94,Jed:Szn:19}. 
For the annealed version of the $q$-voter model, we have
\begin{align}
	\frac{dF_\epsilon}{dc}=&q\left[(1-c)c^{q-1}+c(1-c)^{q-1}\right]-c^q-(1-c)^q\nonumber\\
	&+q\epsilon(1-2c)\left[(1-c)^{q-1}-c^{q-1}\right]\nonumber\\
	&-2\epsilon\left[1-c^q-(1-c)^q\right].
\end{align}
For the disordered phase $c_\text{st}=0.5$ and for the 
fully ordered one $c_\text{st}\in\{0,1\}$, we can determine the stability analytically.  
For $q>1$:
\begin{equation}
F_\epsilon'(0.5)=2^{1-q}(q+2\epsilon-1)-2\epsilon,
\end{equation}
so the disordered state is stable if $\epsilon>\epsilon_1$ and unstable otherwise, where
\begin{equation}
\epsilon_1=\frac{q-1}{2^q-2}.
\end{equation}
For the 
fully ordered states,
\begin{equation}
F_\epsilon'(1)=F_\epsilon'(0)=q\epsilon-1,
\end{equation}
so these states are stable for $\epsilon<\epsilon_2$ and unstable otherwise, where
\begin{equation}
\epsilon_2=\frac{1}{q}.
\end{equation} 
\begin{figure*}[t!]
	\centering
	\subfloat{\label{fig:ppa}\epsfig{file=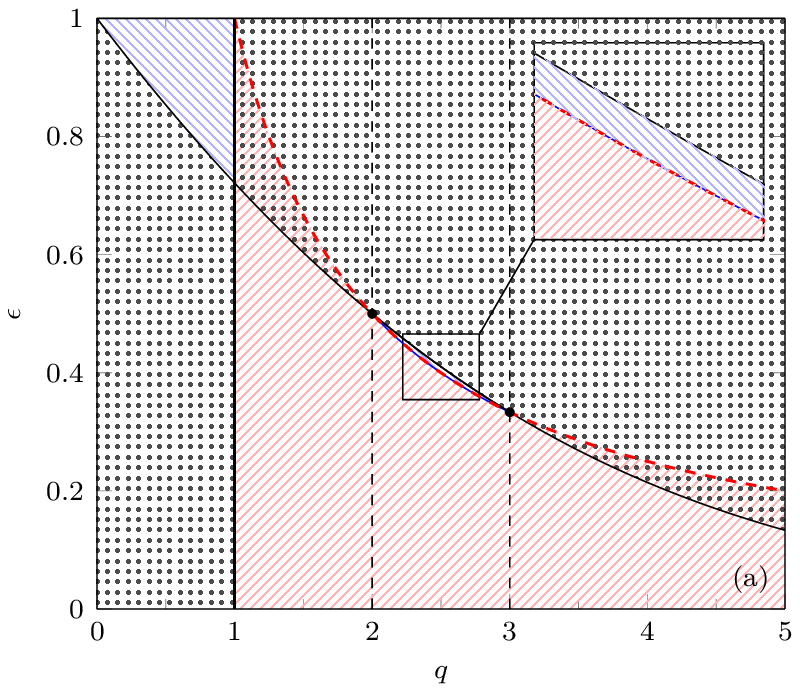,width=8.2cm}}
	\hfill
	\subfloat{\label{fig:ppq}\epsfig{file=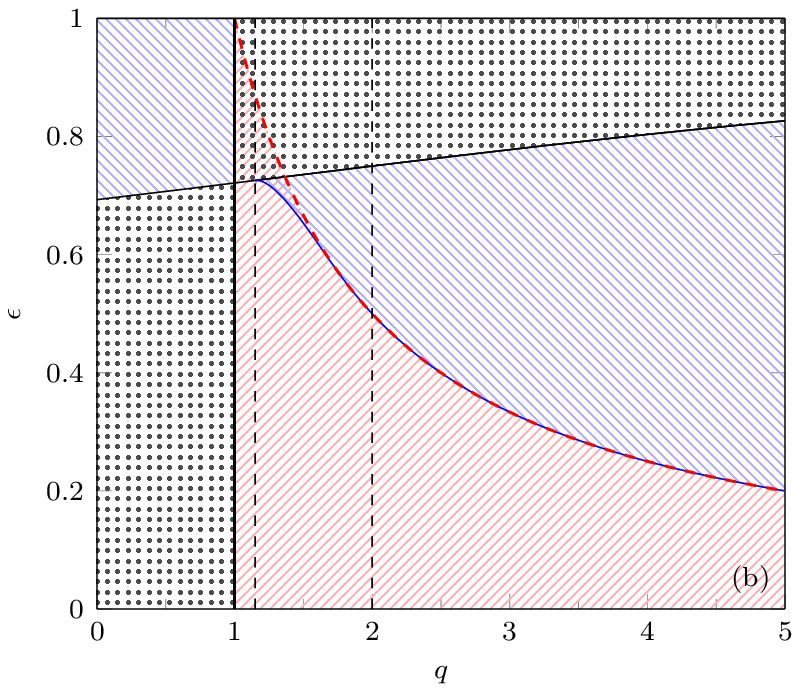,width=8.2cm}}
	\caption{\label{fig:pp-a} Phase diagrams of the nonlinear $q$-voter model under the (a) annealed and (b) quenched approach. The stable fully ordered and active ordered phases are indicated by northeast (red) and northwest (blue) lines, respectively. The stable disordered phase is marked by the dotted area. The thick, vertical, black lines and black dots indicate the voter transitions. The thin, black and thick, dashed, red curves correspond to $\epsilon_1$ and $\epsilon_2$.}
\end{figure*}
On the other hand, when $0<q<1$, the disordered state is stable for $\epsilon<\epsilon_1$ and unstable otherwise, whereas the fully ordered states are unstable for all values of $\epsilon$. 
The stability of the active ordered phase given by Eq.~(\ref{eq:mstatesA}) is determined numerically.
Additionally, as already noted in Ref.\cite{Mor:etal:13}, 
\begin{equation}
\lim_{c_\text{st}\rightarrow 0.5} \epsilon_\text{st}=\frac{q-1}{2^q-2},
\label{eq:limit-a-0.5}
\end{equation}
and
\begin{equation}
\lim_{c_\text{st}\rightarrow 0} \epsilon_\text{st}=\lim_{c_\text{st}\rightarrow 1} \epsilon_\text{st}=\frac{1}{q}
\end{equation}
when $q>1$, so the curve associated with the active ordered states passes through the points where the disordered state and the fully ordered states change their stability, that is, at $\epsilon_1$ and $\epsilon_2$, respectively.
For $0<q<1$, Eq.~(\ref{eq:limit-a-0.5}) still holds, yet the limiting behavior of the concentration approaching the fully ordered phase changes
\begin{equation}
\lim_{c_\text{st}\rightarrow 0} \epsilon_\text{st}=\lim_{c_\text{st}\rightarrow 1} \epsilon_\text{st}=1.
\end{equation}

The phase diagram for the nonlinear $q$-voter model under the annealed approach is presented in  Fig.~\ref{fig:ppa}.
In the diagram, different patterns mark areas where specific phases are stable.
The fully ordered and active ordered phases are indicated by northeast (red) and northwest (blue) lines, respectively. 
The disordered phase is marked by the dotted area. 
The stability borders of disordered and fully ordered phases, $\epsilon_1$ and $\epsilon_2$, are depicted by solid (black) and dashed (red) curves, respectively.
%In the regions $0<q<1$ and $2<q<3$, the above phases are separated by continuous phase transitions; see Figs.~\ref{fig:pda-a} and \ref{fig:pda-d}.
%Outside these areas, an additional mixed phase appears (between $\epsilon_1$ and $\epsilon_2$), in which the fully ordered and disordered phases are both stable, and only discontinuous transitions occur between them; see Figs.~\ref{fig:pda-b} and \ref{fig:pda-f}.
%This mixed phase shrinks towards the points $q=2$ and $q=3$, where it eventually disappears giving rise to voter transitions at the same time; see Figs.~\ref{fig:pda-c} and \ref{fig:pda-e}.
%The voter transitions at the borders of the region $2<q<3$ split into two continuous phase transitions inside of it; see Fig.~\ref{fig:pda-d}.
%Going towards smaller values of $\epsilon$, the first transition at $\epsilon_1$ breaks the up-down symmetry, whereas the second one at $\epsilon_2$ drives the system into one of the fully ordered states already chosen by the symmetry breaking.
%In spatially extended system above one dimension, these transitions should fall into the Ising and the directed percolation universality classes, respectively \cite{Ham:etal:05}.
%In that sense, the voter transition can be understood as a superposition of the above two transitions \cite{Cas:Mun:Pat:09,Ham:etal:05,Vaz:Lop:08,Dro:Fer:Lip:03}.
\begin{figure*}[t!]
	\centering
	\subfloat{\label{fig:pda-a}\epsfig{file=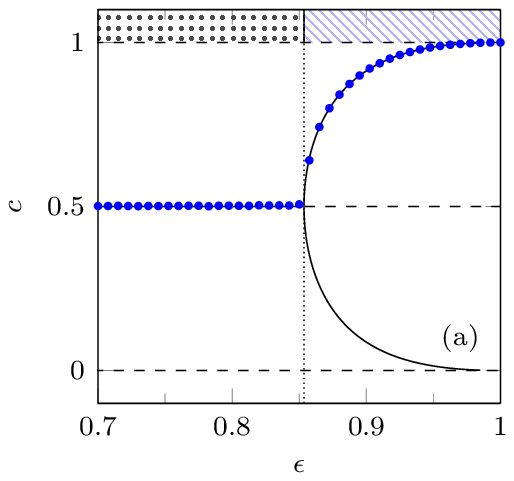,width=5.4cm}}
	\hfill
	\subfloat{\label{fig:pda-b}\epsfig{file=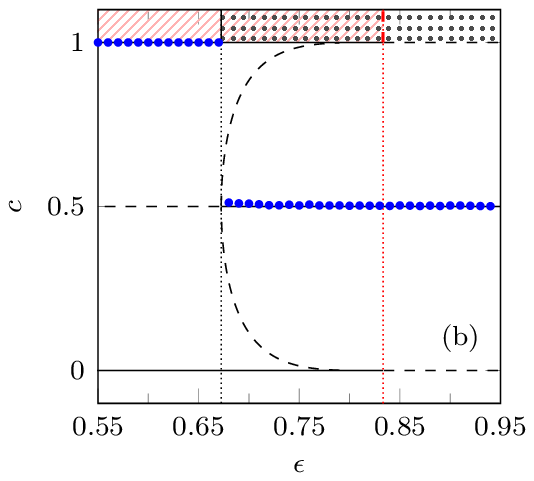,width=5.4cm}}
	\hfill
	\subfloat{\label{fig:pda-c}\epsfig{file=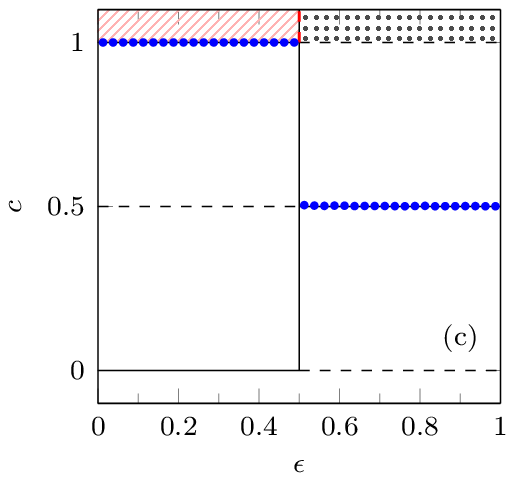,width=5.4cm}}
	\hfill
	\subfloat{\label{fig:pda-d}\epsfig{file=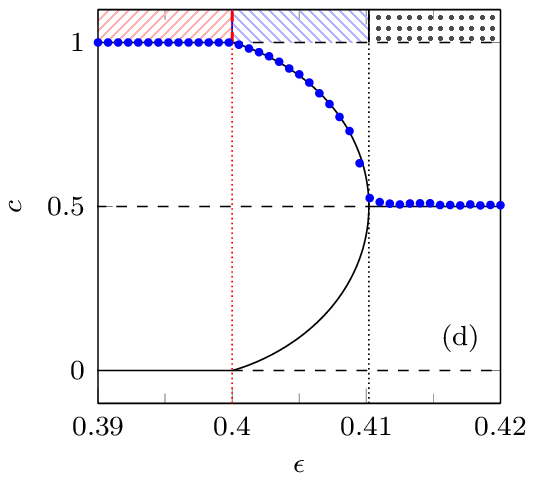,width=5.4cm}}
	\hfill
	\subfloat{\label{fig:pda-e}\epsfig{file=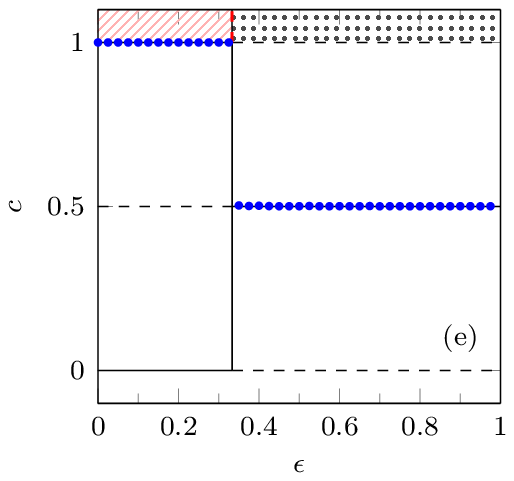,width=5.4cm}}
	\hfill
	\subfloat{\label{fig:pda-f}\epsfig{file=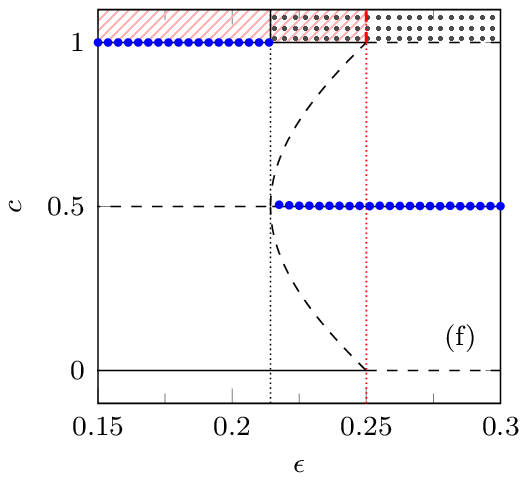,width=5.4cm}}
	\caption{\label{fig:pd-a-0.5} Representative phase diagrams of the nonlinear $q$-voter model under the annealed approach for different values of the influence group size $q$: (a) $0<q<1$, (b) $1<q<2$, (c) $q=2$, (d) $2<q<3$, (e) $q=3$, and (f) $q>3$. Solid and dashed lines represent stable and unstable steady states, respectively.
		The meaning of the marked area above the diagrams accords with the one used in Fig.~\ref{fig:pp-a}.
		Vertical black and red lines correspond to $\epsilon_1$ and $\epsilon_2$, respectively.
		The exact values of $q$ in the plots are as follows: (a) $q=0.5$, (b) $q=1.2$, (d) $q=2.5$, and (f) $q=4$. 
		Dots represent Monte Carlo simulations of the well-mixed system that contains $N=5\cdot10^5$ voters. The results are averaged over 10 runs and collected after 4000 MCS. The initial concentrations amount $c=0.5$.
	}
\end{figure*}

The annealed model behavior can be summarized as follows:
\begin{itemize}
	\item For $0<q<1$, a continuous phase transition occurs at $\epsilon_1$; see Fig.~\ref{fig:pda-a}.
	The fully ordered states are unstable for all values of $\epsilon$.
	The disorder phase shrinks in favor of the active ordered one along with the increasing influence group size $q$.
	The transition disappears in the vicinity of $q=0$ since $\lim_{q\to0}\epsilon_1=1$.
	At the other end of this region, we have $\lim_{q\to1}\epsilon_1=\log^{-1}(4)\approx0.7213$.
	\item For $q=1$, the voter model is obtained.
	\item For $1<q<2$, discontinuous phase transitions between the disordered and the fully ordered states take place; see Fig.~\ref{fig:pda-b}.
	The metastable region, where both these phases are stable creating a mixed phase, lies between $\epsilon_1$ and $\epsilon_2$.
	This area shrinks along with the increasing value of $q$.
	\item For $q=2$, the metastable region disappears completely, and the voter transition at $\epsilon=1/2$ separates the disordered and fully ordered phases; see Fig.~\ref{fig:pda-c}.
	\item For $2<q<3$, which is a typical size of freely forming groups \cite{Rub:etal:13}, two continuous phase transitions occur.
	The first one at $\epsilon_1$ separates the active ordered and the disordered phases. At this point, the up-down symmetry is broken. 
	On the other hand, the second transition at $\epsilon_2$ takes place between the fully ordered and the active ordered phases; see Fig.~\ref{fig:pda-d}.
	In spatially extended system above one dimension, these transitions should fall into the Ising and the directed percolation universality classes, respectively \cite{Ham:etal:05}.
	In that sense, the voter transition can be understood as a superposition of the above two transitions \cite{Cas:Mun:Pat:09,Ham:etal:05,Vaz:Lop:08,Dro:Fer:Lip:03}.
	\item For $q=3$, the region with the active ordered phase disappears.
	 The voter transition, which separates the disordered and the fully ordered phases, emerges at the same time at the point $\epsilon=1/3$; see Fig.~\ref{fig:pda-e}.
	\item For $q>3$, the situation is similar to the one for $1<q<2$; see Fig.~\ref{fig:pda-f}.
	Only discontinuous phase transitions take place between the disordered and the fully ordered phases with a metastable region extending from $\epsilon_1$ to $\epsilon_2$.
	When the parameter $q\to\infty$, the transition disappears, and the disordered phase dominates.
\end{itemize}

Let us notice that by the linearization technique, we obtained the exact value of the point $\epsilon_2$, where the stability of the fully ordered states changes, in contrast to the approximate result from Ref.~\cite{Cas:Mun:Pat:09}.
The differences between these results are mostly noticeable for smaller values of $q$ (especially for $q<2$).
The approximate solution predicts much narrower metastable region for $1<q<2$, and it also indicates the existence of stable fully ordered states for $q<1$, whereas these states are actually unstable then.

\subsection{Quenched model}
\label{sec:quenched-model}
Under the quenched approach, we have two groups of voters: unyielding and vacillating one.
Within each of the groups, we introduce the concentration of voters with opinion $j=1$, and 
we label it $c_0$ and $c_1$ for the unyielding and vacillating voters, respectively.
As a result, the total concentration is a weighted sum of the above concentrations
\begin{equation}
c=(1-\epsilon)c_0+\epsilon c_1.
\end{equation}
Now, the mean-field dynamics is described by two differential equations
\begin{align}
	\frac{dc_0}{dt}&=F_0,\\
	\frac{dc_1}{dt}&=F_1,
\end{align}
where $F_0$ and $F_1$ are appropriate effective forces:
\begin{align}
F_0&=(1-c_0)f_0(c)-c_0f_0(1-c),\\
F_1&=(1-c_1)f_1(c)-c_1f_1(1-c).
\end{align}
As previously, these forces are derived based on the quenched versions of Eq.~(\ref{eg:flipprob}), which are obtained in the following way.
Since the unyielding voters always stick to their opinion in case of divided influence group, we put $\epsilon=0$ in Eq.~(\ref{eg:flipprob}) for them and get
\begin{equation}
f_0(x)=x^q.\label{eq:f0}
\end{equation}
On the other hand, the opinion change is certain for the vacillating voters in a corresponding situation, so $\epsilon=1$ in Eq.~(\ref{eg:flipprob}), which leads to
\begin{equation}
f_1(x)=1-(1-x)^q \label{eq:f1}.
\end{equation}
Using Eqs.~(\ref{eq:f0}) and (\ref{eq:f1}), the effective forces are as follows
\begin{align}
F_0&=(1-c_0)c^q-c_0(1-c)^q, \label{eq:F0}\\
F_1&=(1-c_1)\left[1-(1-c)^q\right]-c_1(1-c^q).\label{eq:F1}
\end{align}
As in the annealed case, steady states $(c_{0\text{st}},c_{1\text{st}})$ are those for which $F_0=F_1=0$. 
Thus, we have $(c_{0\text{st}},c_{1\text{st}})\in\{(0,0), (0.5,0.5),(1,1)\}$ for which Eqs.~(\ref{eq:F0}) and (\ref{eq:F1})  vanish for arbitrary values of $\epsilon$. 
Those  lead to $c_\text{st}\in\{0,0.5,1\}$, respectively.
The rest of the steady states are given by the following dependencies:
\begin{equation}
\epsilon_\text{st}=\frac{\left[(1-c_\text{st})c_\text{st}^q-c_\text{st}(1-c_\text{st})^q\right]\left[c_\text{st}^q+(1-c_\text{st})^q-2\right]}{\left[c_\text{st}^q-(1-c_\text{st})^q\right]\left[c_\text{st}^q+(1-c_\text{st})^q-1\right]},
\label{eq:qcstotal}
\end{equation}
\begin{equation}
c_{0\text{st}}=\frac{c_\text{st}^q}{c_\text{st}^q+(1-c_\text{st})^q},
\label{eq:qc0s}
\end{equation}
and
\begin{equation}
c_{1\text{st}}=\frac{1-(1-c_\text{st})^q}{2-c_\text{st}^q-(1-c_\text{st})^q}.
\label{eq:qc1s}
\end{equation}
The stability of a steady state is established based on the determinant and trace of the Jacobian matrix evaluated at this point
\begin{equation}
\mathbf{J}(c_{0\text{st}},c_{1\text{st}})=\begin{bmatrix}
\frac{\partial F_0}{\partial c_0} & \frac{\partial F_0}{\partial c_1}\\
\frac{\partial F_1}{\partial c_0} & \frac{\partial F_1}{\partial c_1}
\end{bmatrix}_{(c_0,c_1)=(c_{0\text{st}},c_{1\text{st}})},
\end{equation}
where
\begin{align}
	\frac{\partial F_0}{\partial c_0}=&q(1-\epsilon)\left[(1-c_0)c^{q-1}+c_0(1-c)^{q-1}\right]\nonumber\\
	&-c^q-(1-c)^q,\\
	\frac{\partial F_0}{\partial c_1}=&q\epsilon\left[(1-c_0)c^{q-1} + c_0(1-c)^{q-1}\right],\\
	\frac{\partial F_1}{\partial c_0}=&q(1-\epsilon)\left[(1-c_1)(1-c)^{q-1}+c_1c^{q-1}\right],\\
	\frac{\partial F_1}{\partial c_1}=&q\epsilon\left[(1-c_1)(1-c)^{q-1}+c_1c^{q-1}\right]\nonumber\\
	&+c^q+(1-c)^q-2.
\end{align}
The state is stable if $\det\left[\mathbf{J}(c_{0\text{st}},c_{1\text{st}})\right]>0$ and $\tr\left[\mathbf{J}(c_{0\text{st}},c_{1\text{st}})\right]<0$ \cite{Str:94}.
Once again, for the disordered phase $c_\text{st}=0.5$ and for the fully ordered  states $c_\text{st}\in\{0,1\}$, we are able to determine the stability analytically.
For $q>1$, we have
\begin{align}
	\det\left[\mathbf{J}(0.5,0.5)\right]&=4^{1-q}[q\epsilon(2^q-2)-(q-1)(2^q-1)],\\
	\tr\left[\mathbf{J}(0.5,0.5)\right]&=2^{1-q}q-2.
\end{align}
Thus, the disordered phase is stable for $\epsilon>\epsilon_1$ and unstable otherwise, where
\begin{equation}
\epsilon_1=\frac{(q-1)(2^q-1)}{q(2^q-2)},
\end{equation}
so this stability boundary differs from the one in the annealed model.
However, for the fully ordered states,  we get the same result as in the annealed case since
\begin{align}
\det\left[\mathbf{J}(0,0)\right]&=\det\left[\mathbf{J}(1,1)\right]=1-q\epsilon,\\
\tr\left[\mathbf{J}(0,0)\right]&=\tr\left[\mathbf{J}(1,1)\right]=q\epsilon-2,
\end{align}
which leads to the conclusion that the fully ordered states are stable if $\epsilon<\epsilon_2$ and unstable otherwise, where
\begin{equation}
	\epsilon_2=\frac{1}{q}.
\end{equation}
\begin{figure*}[t]
	\centering
	\subfloat{\label{fig:pdq-a}\epsfig{file=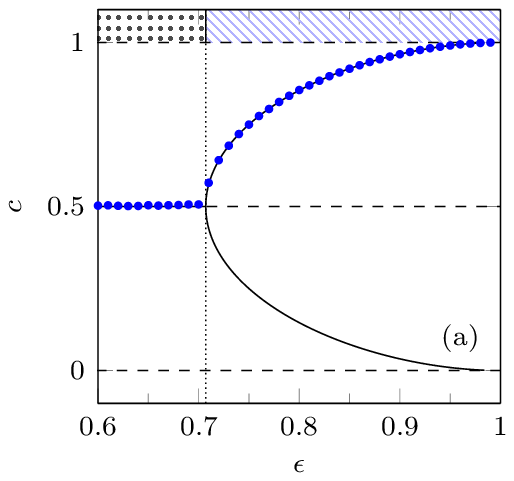,width=5.4cm}}
	\hfill
	\subfloat{\label{fig:pdq-b}\epsfig{file=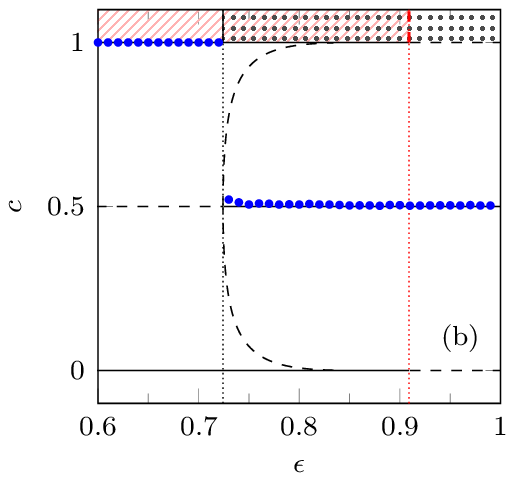,width=5.4cm}}
	\hfill
	\subfloat{\label{fig:pdq-c}\epsfig{file=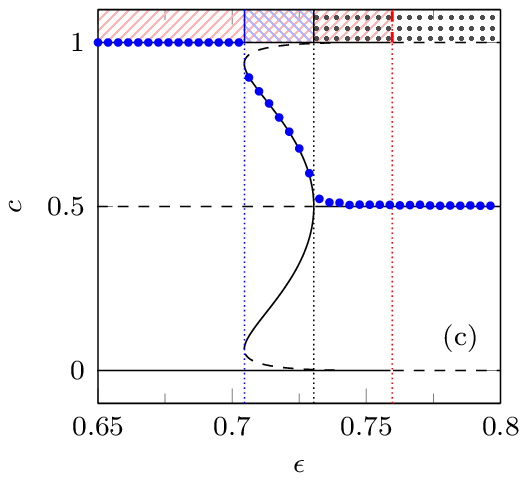,width=5.4cm}}
	\hfill
	\subfloat{\label{fig:pdq-d}\epsfig{file=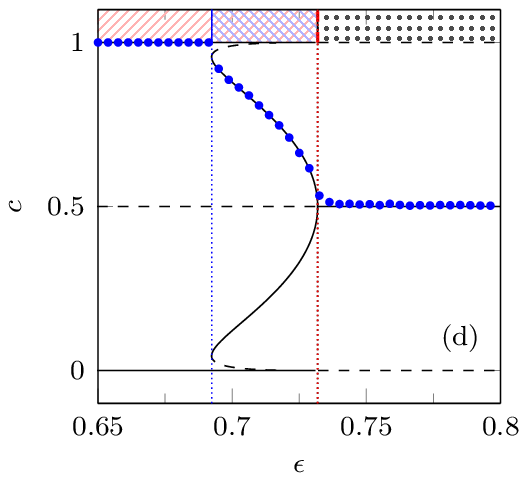,width=5.4cm}}
	\hfill
	\subfloat{\label{fig:pdq-e}\epsfig{file=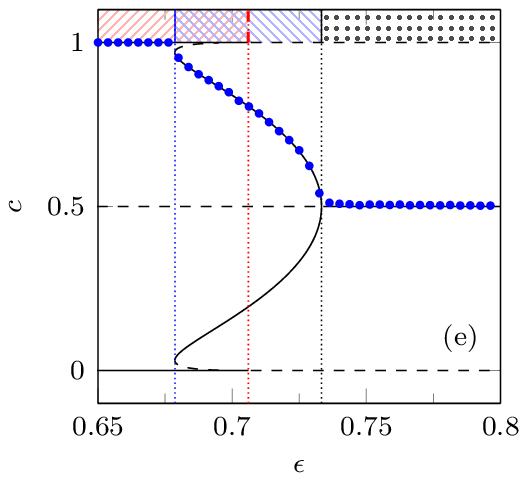,width=5.4cm}}
	\hfill
	\subfloat{\label{fig:pdq-f}\epsfig{file=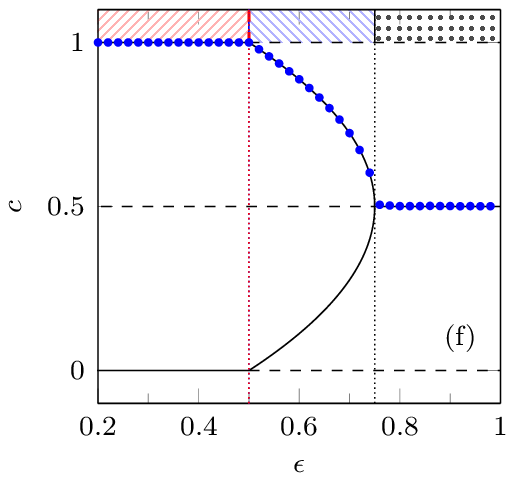,width=5.4cm}}
	\caption{\label{fig:pd-q-0.5} Representative phase diagrams of the nonlinear $q$-voter model under the quenched approach for different values of the influence group size $q$: (a) $0<q<1$, (b) $1<q<q^*$, (c) $q^*<q<\bar{q}$, (d) $q=\bar{q}$, (e) $\bar{q}<q<2$, and (f) $q\geq2$. 
		At the point $q^*\approx1.1493$, a new mixed phase emerges with both stable fully ordered and active ordered states, whereas at $\bar{q}\approx1.3664$, we have $\epsilon_1=\epsilon_2$.
		Solid and dashed lines represent stable and unstable steady states, respectively.
		The meaning of the marked area above the diagrams accords with the one used in Fig.~\ref{fig:pp-a}.
		Vertical black and red lines correspond to $\epsilon_1$ and $\epsilon_2$, respectively.
		The exact values of $q$ in the plots are as follows: (a) $q=0.5$, (b) $q=1.1$, (c) $q=\bar{q}-0.05\approx1.3164$, (e) $q=\bar{q}+0.05\approx1.4164$, and (f) $q=2$. 
		Dots represent Monte Carlo simulations of the well-mixed system that contains $N=5\cdot10^5$ voters. The results are averaged over 10 runs and collected after $10^4$ MCS. The initial concentrations amount $c=0.5$. }
\end{figure*}
When $0<q<1$, the disordered phase is stable for $\epsilon<\epsilon_1$ and unstable otherwise, whereas the fully ordered states are unstable for all values of $\epsilon$.
The stability of the active ordered states given by Eqs.~(\ref{eq:qcstotal})-(\ref{eq:qc1s}) is determined numerically.
As in the annealed model, the curve associated with these states passes through the points $\epsilon_1$ and $\epsilon_2$ when $q>1$ since then we have
\begin{equation}
\lim_{c_\text{st}\rightarrow 0.5} \epsilon_\text{st}=\frac{(q-1)(2^q-1)}{q(2^q-2)},
\end{equation}
and
\begin{equation}
\lim_{c_\text{st}\rightarrow 0} \epsilon_\text{st}=\lim_{c_\text{st}\rightarrow 1} \epsilon_\text{st}=\frac{1}{q}.
\end{equation}
For $0<q<1$, only the limits for the  fully ordered states change, and as previously we have
\begin{equation}
\lim_{c_\text{st}\rightarrow 0} \epsilon_\text{st}=\lim_{c_\text{st}\rightarrow 1} \epsilon_\text{st}=1.
\end{equation}

Figure~\ref{fig:ppq} illustrates the phase diagram for the nonlinear $q$-voter model under the quenched approach.
As seen, there are several striking differences between the models with annealed and quenched randomness.
First of all, the stability boundary of the disordered phase $\epsilon_1$ is an increasing function within the quenched approach in contrast to the annealed case, where it decreases with the parameter $q$.
As a result, the models exhibit different limiting behaviors.
In the vicinity of $q=0$,  the phase transition in the model with annealed randomness disappears, whereas in the model with quenched randomness, it persists  with the transition point approaching $\lim_{q\to0}\epsilon_1=\log(2)\approx0.693$.
On the other hand, when $q\to\infty$, neither of the models exhibit a phase transition, and in both cases only one phase survives -- the disordered one in the annealed case and the active ordered one in the quenched case.
Another feature of the quenched model is the appearance of a new mixed phase unobserved neither in the original $q$-voter model \cite{Cas:Mun:Pat:09} nor in its extended version with a threshold \cite{Vie:Ant:18}.
In this phase, both the fully ordered and active ordered states are stable, whereas the disordered state is unstable, see Figs.~\ref{fig:pdq-c}-\ref{fig:pdq-e}.
It first emerges at the point $q^*\approx1.1493$, where the subcritical pitchfork bifurcation turns into supercritical one (compare Figs.~\ref{fig:pdq-b} and \ref{fig:pdq-c}, for which $q$ is a little bit smaller and larger than $q^*$, respectively).
This specific point can be derived from Eq.~(\ref{eq:qcstotal}) by applying the following condition
\begin{equation}
	\lim_{c_\text{st}\to 0.5}\frac{d^2\epsilon_\text{st}}{dc^2_\text{st}}=0
\end{equation}
since then the concavity of the function $\epsilon_\text{st}(c_\text{st})$ near $c_\text{st}=0.5$ changes.
The region with this new mixed phase extends up to $q=2$, where the points of saddle node bifurcations hit the fully ordered borders at $\epsilon_2=1/2$; see Fig.~\ref{fig:pdq-f}.
Within this area, another characteristic point $\bar{q}\approx1.3664$ can be identified at which the stability of the disordered phase and the fully ordered one changes at the same time, that is, $\epsilon_1=\epsilon_2$.
At this point, only our new mixed phase separates the disordered and the fully ordered phases, see Fig.~\ref{fig:pdq-d}.
Below it, the supercritical pitchfork bifurcation associated with the up-down symmetry breaking occurs inside the hysteresis loop (see Fig.~\ref{fig:pdq-c}), whereas above it, the symmetry breaks first, and afterwards the system falls into one of the fully ordered states by a discontinuous jump (see Fig.~\ref{fig:pdq-e}).
For $q\geq2$, the system undergoes two continuous phase transitions.
One breaks the symmetry; the other puts the system into the fully ordered phase, see Fig.~\ref{fig:pdq-f}.

The quenched model behavior can be summarized as follows:
\begin{itemize}
	\item For $0<q<1$, the continuous phase transition takes place between the disordered and the active ordered phases; see Fig.~\ref{fig:pdq-a}.
	The transition point $\epsilon_1$ increases along with the parameter $q$ starting from the value $\lim_{q\to0}\epsilon_1=\log(2)\approx0.6931$.
	Thus, the transition persists even close to $q=0$; see Fig.~\ref{fig:ppq}.
	At the other end of this region, we have $\lim_{q\to1}\epsilon_1=\log^{-1}(4)\approx0.7213$.
	\item For $q=1$, the voter model is obtained.
	\item For $1<q<q^*$, the discontinuous phase transitions occur between the disordered and the fully ordered phases; see Fig.~\ref{fig:pdq-b}. The metastability associated with the mixed phase is confined to the region between $\epsilon_1$ and $\epsilon_2$.
	This metastable area shrinks along with the increasing value of $q$.
	\item For $q^*<q<\bar{q}$, a new mixed phase that combines the fully ordered and the active ordered phases appears at the point $q^*\approx1.1493$.
	Moreover, the already known mixed phase with the disordered and the fully ordered phases still exists up to the point $\bar{q}\approx1.3664$.
	Consequently, this is the area where a continuous phase transition breaks the up-down symmetry inside the hysteresis loop; see Fig.~\ref{fig:pdq-c}.
	\item For $q=\bar{q}$, the mixed phase with the disordered and the fully ordered phases disappears since $\epsilon_1=\epsilon_2$ at this point.
	As a result, the only mixed phase that remains combines the fully ordered and the active ordered phases; see Fig.~\ref{fig:pdq-d}.
	\item For $\bar{q}<q<2$, the active ordered phase appears right after crossing $\bar{q}$, and it extends.
	On the other hand, the mixed phase shrinks and eventually disappears at the point $q=2$, in which the saddle node bifurcations reach the fully ordered borders.
	In this area, a continuous transition precedes the hysteresis loop between the fully ordered and the active ordered states; see Fig.~\ref{fig:pdq-e}.
	\item For $q\geq2$, only continuous phase transitions are possible; see Fig.~\ref{fig:pdq-f}.
	The transition points, $\epsilon_1$ and $\epsilon_2$, tend to 1 and 0, respectively, as the parameter $q\to\infty$.
	Thus, the active ordered phase expands and eventually fills the entire diagram.
\end{itemize}

\section{Conclusions}
We compared two versions of the nonlinear $q$-voter model -- with annealed and quenched randomness -- in order to find out how different disorders impact phase transitions exhibited by the system.
Mean-field analysis used in the study revealed several striking differences between these two approaches to modeling voters' behavior.
First of all, quenched randomness eliminates most of the discontinuous phase transitions present in the annealed model.
In the cases where the transitions were originally continuous, on the other hand, it extends the active ordered phase.
Similar situations have already been reported in other models of opinion dynamics, including Galams's model \cite{Sta:Mar:04} or the $q$-voter model with independence \cite{Jed:Szn:17}, and in models of equilibrium statistical mechanics \cite{Aiz:Weh:89,Hui:Ber:89,Cha:Ber:98}.
Furthermore, the modified model exhibits a mixed phase, which combines the fully ordered and active ordered states, unobserved neither in the original model \cite{Cas:Mun:Pat:09} nor in its extended threshold modification \cite{Vie:Ant:18}.
The existence of this phase is connected with the specific combinations of phase transitions in which the up-down symmetry breaks inside the hysteresis loop, or the symmetry breaks first, and afterwards a discontinuous transition to one of the fully ordered states occurs.
Such sequences of transitions have not been reported in any of the $q$-voter model modifications.
Finally, the limiting behavior of the system connected with the parameter $q$ is different.
When $q\to0$, the transitions in the annealed model disappears.
In the quenched version, however, the transitions are more robust, and they survive even in close vicinity of $q=0$.
On the other hand, when $q\to\infty$, one phase dominates the entire phase space in both models. The disordered one in the case of annealing and the active ordered one in the case of quenching.

These differences and new phenomena induced by quenched randomness should encourage further research on models of opinion dynamics with different types of disorders.
Such studies are not only psychologically grounded, but they also help to advance our knowledge about phase transitions.  
Especially, the analysis of models that already exhibit discontinuous phase transitions may be interesting in that context, like recently introduced generalizations of the $q$-voter model with different types of nonconfomity \cite{Abr:Paw:Szn:19, Nyc:etal:18}, the extension of Watts' model with independent behavior \cite{Now:Szn:19}, or the nonlinear $q$-voter model with a threshold \cite{Vie:Ant:18}.
Going beyond the standard mean-field theory and introducing different topologies to the problem is another challenge that should be faced in the future \cite{Gle:13,Mor:etal:13, Jed:17,Per:etal:18}.
%Regular lattices and complex networks are here to our disposal.

\section*{Acknowledgments}
This work was created as a result of the research projects nos. 2018/28/T/ST2/00223, 2016/23/N/ST2/00729, and 2016/21/B/HS6/01256 financed from the funds of the National Science Center (NCN, Poland).

\bibliography{JedSzn_literature}

\end{document}